\documentclass[newstyle,article]{rmaa}

\newcommand{\tsq}{\hbox{$t^2$}}
\newcommand{\ii}{\'{\i}}
\newcommand{\micr}{\,$\mu$m\/}
\newcommand{\map}{\hbox{{\sc mappings i}c}}

\newcommand{\cmc}{\hbox{\,${\rm cm^{-3}}$}}

\newcommand{\pc}{\hbox{\,pc}}
\newcommand{\mud}{\hbox{$\mu_D$}}

\newcommand{\thetac}{\hbox{$\theta^1$\,C~Ori}}

\newcommand{\nii}{\hbox{[N\,{\sc ii}]}}

\newcommand{\niiw}{\hbox{[N\,{\sc ii}]$\lambda $6583}}
\newcommand{\niitw}{\hbox{[N\,{\sc ii}]$\lambda $5755}}
\newcommand{\oiii}{\hbox{[O\,{\sc iii}]}}
\newcommand{\oiiiw}{\hbox{[O\,{\sc iii}]$\lambda $5007}}
\newcommand{\oiiitw}{\hbox{[O\,{\sc iii}]$\lambda $4363}}
\newcommand{\oii}{\hbox{[O\,{\sc ii}]}}
\newcommand{\oiir}{\hbox{O\,{\sc ii}}}
\newcommand{\oiiw}{\hbox{[O\,{\sc ii}]$\lambda\lambda $3727}}

\newcommand{\roiii}{\hbox{R$_{{\rm OIII}}$}}
\newcommand{\rnii}{\hbox{R$_{{\rm NII}}$}}

\newcommand{\ha}{\hbox{H$\alpha$}}

\newcommand{\hb}{\hbox{H$\beta$}}
\newcommand{\hbw}{\hbox{H$\beta$\,$\lambda $4861}}

\newcommand{\hii}{\hbox{H\,{\sc ii}}}
\newcommand{\sii}{\hbox{[S\,{\sc ii}]}}

\title{Density gradients and internal dust in the Orion nebula}
\author{L. Binette\altaffilmark{1}, D. I. Gonz\'alez-G\'omez\altaffilmark{1} and
Y. D. Mayya\altaffilmark{2}}

\altaffiltext{1}{Instituto de Astronom\'\i a, UNAM, M\'exico.}
\altaffiltext{2}{INAOE, Tonantzintla, M\'exico.}

\fulladdresses{
\item L. Binette  and D. I Gonz\'alez-G\'omez:  Instituto de Astronom\ii
a, UNAM, Apartado Postal 70-264, 04510 D.F., M\'exico  
(dulce, binette@astroscu.unam.mx). 

\item Y. D. Mayya: Instituto Nacional de Astrof\ii sica, Optica y
Electr\'onica, Apartado Postal 216, 7200 Puebla, Pue., M\'exico.
}

\shortauthor{Binette et al.}
\shorttitle{Density gradient in Orion}
\ReceivedDate{2002 July } 
\AcceptedDate{2002 September 13} 


\abstract{
The ionization structure of the Orion nebula can be described as a
skin-like ionization structure on the surface of a dense cloud. We
propose that a steep density stratification, increasing as a powerlaw
($n \propto x^{-2}$) function of distance $x$ from the ionization front,
exhibits properties which agree with our long-slit spectrum of the
Orion nebula. For instance, there exist a unicity relation between
both the \hb\ surface brightness or the ionization front \sii\
density, and the scale $L$ of the powerlaw, where $L$ is the distance
between the ionization front and the onset of the density 
near the exciting star.  Internal dust is required to obtain a
simultaneous acceptable fit of both the \sii\ density and then \hb\ suface
brightness observations.  Nebular models containing small dust grains
provide a better fit than large grains. The line ratio gradients
observed along the slit are qualitatively reproduced by our density
stratified models assuming a stellar temperature of 38000
K. Collisional deexcitation appears to be responsible for half of the
gradient observed in the \niitw/\niiw\ temperature sensitive ratio. We
propose that the empirical relationship found by Wen \& O'dell between
density and stellar distance may possibly be caused by a power-law
density stratification.  }

\nonstopmode

\begin{document}

\maketitle
\setcounter{footnote}{0}
\section{Introduction}
\label{sec:intro}

It was recognized early on by M\"unch (1958) and Wurm (1961) that the
Orion nebula is better described as a photoionized skin on the surface
of a dense cloud than by a uniformly filled gas structure.
Furthermore, kinematics of the ionized gas (see review by O'Dell 1994)
indicates the gas is flowing away from the dense ionizing front on the
surface of the molecular cloud.  Although earlier studies proposed an
exponential function to describe the density behaviour from such a
flow (Tenorio-Tagle 1979; Yorke 1986), more recent theoretical and
observational works (Franco, Tenorio-Tagle \& Bodenheimer 1989, 1990;
Franco et al 2000a,b) indicate that expanding \hii\ regions are well
described by a power law density stratification $n \propto x^{-\beta}$
with exponents larger than $\beta =1.5$. This is qualitatively
consistent with the results obtained from hydrodynamic calculations of
photoevaporative flows off proplyd disks by Richling \& Yorke (2000)
or off spherical surfaces by Bertoldi (1989) and Bertoldi \& Draine
(1996).  The current paper looks at how the observed gradient of the
\sii\ density in Orion can be accounted for using a simple powerlaw
density model and how well the surface brightness compares with the
long-slit observations presented here. The aim is to explore the
benefits of a powerlaw density stratification in the context of a
determination of the shape of the ionization front (hereafter IF)
using a surface brightness map in \ha, a task first performed by Wen
\& O'Dell (1995) for the two cases of   isochoric and exponential
density behaviour.

\section{Observations and data reduction}

\begin{figure} 
  \begin{center} \leavevmode
  \includegraphics[width=\columnwidth]{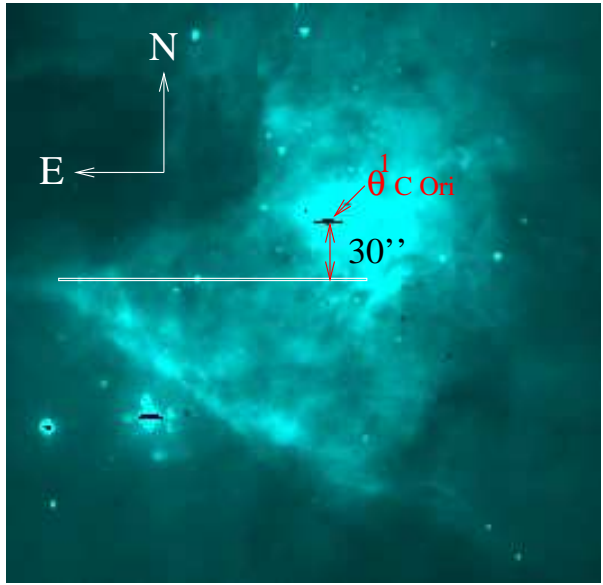}
\caption{Image illustrating the position of our long-slit on an R-band
image of Orion. The position of the exciting star \thetac\ (whose
image is saturated) is indicated. The R-band filter encompasses \ha\
and the red \nii\ lines which account for most of the nebular emission
visible on the image. } \label{fig:slit}
\end{center}
\end{figure}

We have taken long-slit spectra of the central region of Orion at the
Observatorio Astrof\ii sico Guillermo Haro, Cananea, on the 21st of
December 1998 using a $1024\times1024$ TEK CCD detector mounted on a
Boller \& Chivens spectrograph.  We used a grating of 300 l/mm and
covered the wavelength range 3600--5200 \AA, 5500--7100 \AA, and
8000--9600 \AA\ in three settings.  The corresponding exposure times
were 90, 60 and 300 seconds, respectively. A slit width of 2.5\arcsec\
was used. The effective spectral resolution was $\approx 6$ \AA. The
slit was aligned West-East and off-centered by 30\arcsec\ to the South
of \thetac\ as illustrated in Fig.~\ref{fig:slit}. Arc spectra were taken as
well as standard star exposures. Data reduction (bias subtraction,
flat fielding, wavelength and flux calibration) was performed with
IRAF. Due to non-photometric conditions and partial cloud coverage, we
could not derive a reliable absolute flux calibration. We hence used
the calibration of Rodr\'\i guez (1999), who had also observed with
the slit aligned East-West with her regions A-1 and A-3 common to
ours. We tied our H$\beta$ fluxes to the mean surface brightness
within her region A-1. In any event, the adopted Rodr\'\i guez surface
brightness value is only 25\% higher than our initial
calibration. Line fluxes were reddening corrected using the Balmer
decrement and the galactic reddening law (Seaton 1979).  Further
details on the data can be found in Gonz\'alez-G\'omez (1999).

\section{The photoionization models} \label{param}

We adopt a similar approach to that of Binette \& Raga (1990,
hereafter BR90) who studied the properties of powerlaw densities in
photoionized slabs (see also Williams 1992). However, we adapted their
simple slab geometry to the more appropriate spherical geometry for reasons
given below in Sect.\ref{geo}. Another paper which  considers powerlaw
density stratification in photoionization models is that of 
Sankrit \& Hester (2000) (see also Hester et al. 1996). 

The emission line measures in Orion are dominated by the gas near the
IF as a result of the strong density gradient towards it. In the work
of Wen \& O'Dell (1995, hereafter WO95), the authors approximated the
front structure as gas segments (or columns) aligned towards {\it our}
line of sight with a projected area on the sky of $2\arcsec \times
2$\arcsec.  In their final best model, the densities within the
segments declined exponentially towards the observer rather than
radially towards \thetac. Using the observed distribution of \ha\
surface brightness and \sii\ gas density, WO95 were able to solve for
the IF distance (along our line of sight) for the whole nebula. Such a
distribution of distances is termed the ``shape'' of the IF.  The gas
segments of their model did not extend up to the exciting star, a
characteristic also shared by the isobaric slab model of Baldwin et
al. (1991, hereafter BF91). In such a scheme, the segments are
independent and a density must be specified for each segment. In this
Paper, we will explore density distributions which are exclusively
function of radius from \thetac\ without assuming any significant gas
cavity near \thetac.

\begin{figure} 
  \begin{center} \leavevmode
  \includegraphics[width=0.7\columnwidth,angle=-90]{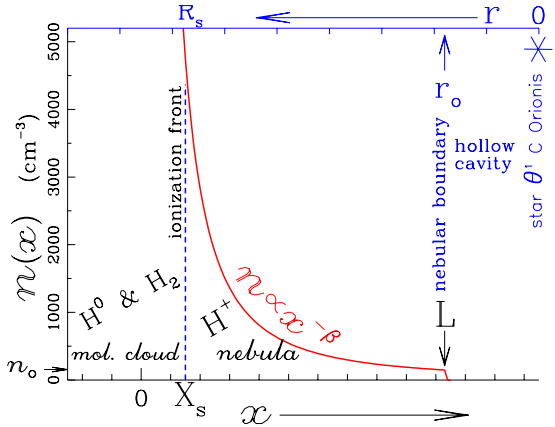}
\caption{Diagram illustrating qualitatively  our   
powerlaw density stratification as a function of $x$. The origin of
the spatial $x$ axis lies at a depth $L$ from the nebular boundary and
the density at that position (at $x=L$) is labelled $n_o$ (see
Equation \ref{eq:pow}).  Note that $L$ will be allowed to vary with
angular direction from \thetac. The top axis is the radial distance
$r$ from the star \thetac, with $r$ increasing towards the left. The
IF (or inner Str\"omgren boundary) is situated at $X_s$ (or
equivalently at $R_s$). The electronic density inferred from the \sii\
doublet corresponds to a density near the IF. The hollow gas cavity
surrounding the star has a size $r_o$.  In our models, the adopted
value for $r_o$ is much smaller than depicted above.}
\label{fig:struc}
\end{center}
\end{figure}

\subsection{The radial density gradient} \label{law}

We adopt a similar notation to BR90 in which the density profile (with
an origin lying inside the dense cloud) is described as follows:

\begin{equation}
\begin{array}{lr}
n(x)=n_o\left({L\over x}\right)^\beta\; ;x\le L\, , \\

n(x)=0\; \; x>L\, , \label{eq:pow}
\end{array} 
\end{equation}
where $n(x)$ is the gas density, $\beta$ the powerlaw index ($n
\propto x^{-\beta}$), $n_o$ the boundary density of the nebula towards
the exciting star, and $x$ the variable representing the distance from
the inner density discontinuity at $x=0$.  Note the reversal of axis
relative to traditional studies: the ionizing radiation is entering at
the cloud boundary $x=L$ and is absorbed inward up to the inner
Str\"omgren boundary at $X_s$ (see Fig.~\ref{fig:struc}). The steep
density gradient near the origin ensures that the photons are absorbed
before the density discontinuity, hence $X_s \ga 0$.

In the case of Orion, the spherical geometry is necessary in order to
take into account the geometrical dilution of the ionizing radiation
across the radial density structure.  One can define the variable,
$r$, as the distance from the central ionizing star; that is $r = L -
x + r_o$ where $r_o$ is the radius of the central hollow cavity and
$n_o$ the gas density at $r_o$.  When the ionizing photon luminosity
$Q_H$ is sufficiently high (or the density $n_o$ low), the Str\"omgren
radius, $R_s = L - X_s + r_o$, is near the origin of the $x$ axis, at
a distance $\simeq L + r_o$ from the exciting star (since $X_s$ is
negligible).  This is referred by BR90 as the ``strong gradient''
regime. In the converse case, for very low ionizing photon
luminosities $Q_H$ (or for high boundary densities $n_o$), $R_s$
shrinks ($X_s$ becomes non-negligible) and becomes of order $r_o + $
small fraction of $L$, a condition which BR90 labeled ``weak
gradient'' because the nebula's thickness is smaller than the density
gradient's scale $L$.  All the quantities introduced above are illustrated in
Fig.~\ref{fig:struc}.

The ``strong gradient'' regime is the appropriate one for Orion since
the nebula is traditionally approximated by a skin-like structure on
the face of a dense molecular cloud (BR90; O'Dell 1994).  One
advantage of the powerlaw stratification when the regime of ``strong
gradient'' prevails is that we know before hand the approximate radius
of the IF ($\la L + r_o $).  In the case of an exponential
stratification, which is much shallower than a powerlaw, it requires
solving for the ionizing radiation transfer inside the nebula in order
to predict the IF depth.

For definiteness, we will consider the case of a powerlaw index $\beta = 2$.
Very similar nebular line ratios of \oii, \oiii, etc. were
obtained with indices of $\beta=1.5$ and 1.0, provided we adjusted $n_o$
and $L$ in such a way that both the ionized depth $R_s$ and the \sii\
density remained the same as that obtained with the $\beta =2$
case. We consider that our results do not depend on the particular
choice of $\beta$ made here.

We found no justification for having an arbitrary large region devoid
of gas around \thetac\ and, for definiteness, have set $r_o = 0.007 {\rm pc}
\la 0.05 R_s$, a size so small that its precise value has no impact on
the calculations.

\subsection{The geometry} \label{geo}

The geometry of the IF is  highly structured and complex
(c.f. WO95) and cannot be modeled directly by a code like
\map\ which is designed to handle simple nebular geometries. 
Our approach has been to compute a sequence of spherical models of
different scale $L$ (c.f. Equation \ref{eq:pow}) in order to simulate
small areas of the IF located at different distances from the exciting
star. This piecemeal approach (a different model for each position
along the slit) allowed us to reconstruct the IF contribution through
each aperture element. To account for the projection of each
slit aperture element onto the calculated Str\"omgren spheres,  
we carried out in \map\ the integral of line emissivities over spatial
limits corresponding to the appropriate aperture projection onto the 
calculated spheres.

Our aim is to model only the background IF (the component B in BF91) which
is much brighter than the foreground IF (the component A in BF91). We
therefore halved the fluxes computed by \map\ to reduce the geometry
to that of an hemisphere rather than a sphere. Furthermore, to remove
the small contribution due to component A, we multiplied the {\it
observed} \hb\ surface brightness by $\case{4}{5}$, the same value as
in BF91.

To sum up, the different IF positions covered by the slit are
associated to a sequence of models which differ only by their value of
the scale $L$. Since our slit was offset to the South by 30\arcsec\
from \thetac\ (extending mostly eastward), it did not sample gas
components which overlaid along common radial lines from the exciting
star.

\subsection{Metallicity, dust and $Q_H$} \label{dust}

Dust mixed with nebular gas, a possibility studied by  BF91, can  play an
important role in the structure of the Orion nebula. We  define
\mud\ as the amount of dust (by mass) present in the gas in units of the
solar neighborhood value. The code \map\ includes the effects of
internal dust as described in Binette et~al. (1993). We will consider
two extinction curves which present quite a different behavior in the
far-UV, both calculated by Martin \& Rouleau (1991): $a)$ the $orion$
curve\footnote{The $orion$ extinction curve is certainly indicated for 
the cold gas in front of the Orion nebula (inside the ``Lid'') 
but is not necessarily appropriate for the warm ionized gas.} 
used by BF91 is characterized by a population of relatively large grain
sizes covering the range 0.03--0.25\micr, $b)$ the $noiro$ curve, 
first introduced by Magris, Binette \& Martin (1993), is characterized
by a population of small grain sizes covering the range
0.005--0.03\micr.  A possible justification is that photoerosion,
which  is significant in   photoionized gas,  
could have the effect of reducing the
mean grain size to that characterizing  the $noiro$
curve. Interestingly, a distribution  of small grains may 
also play an important role in resolving the temperature
problem in \hii\ regions (Stasi\'nska \& Szczerba 2001). Note
that   the solar neighborhood extinction curve as calculated by Martin
\& Rouleau (1991)  is composed of grain sizes covering the full range  
0.005--0.25\micr\ (i.e. $ism = orion + noiro$). The advantage of using
the $noiro$ and the $orion$ curves is that they provide a simple
characterization of fairly opposite cases regarding  extinction
properties in the UV.

We adopt the same gas phase abundances as those determined by BF91
(mean metallicity at 0.7$Z\sun$) and the stellar atmosphere models of
Hummer \& Mihalas (1970) with $T_{eff} = 38000$\,K, using the scheme
proposed by Shields \& Searle (1978) to interpolate in effective
temperature and metallicity between atmosphere models. The choice of
temperature is discussed in Sect.~\ref{lines}.

As argued by WO95, the observed recombination lines surface brightness
includes a reflected back-component due to scattering from the
background dusty molecular cloud. To correct for this, as in WO95 we
multiplied the observed \hb\ surface brightnesses by $\case{2}{3}$ and
adopted a reduced stellar photon luminosity of $Q_H=10^{49}$ photons
${\rm s}^{\rm -1}$ in the models (see discussion in WO95).

\begin{figure} 
  \begin{center} \leavevmode
  \includegraphics[width=\columnwidth]{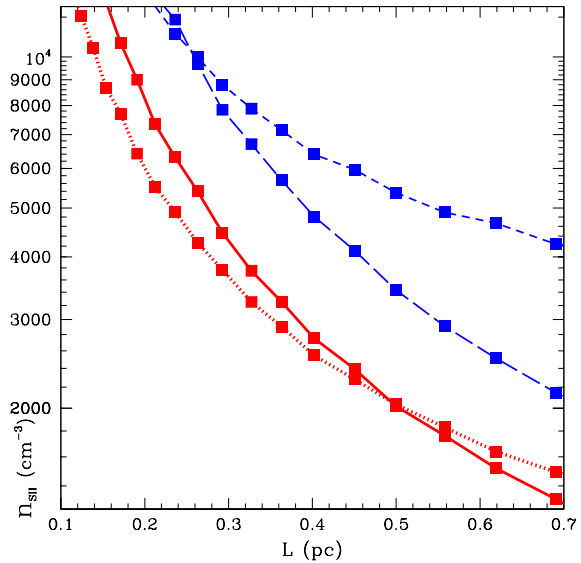}
\caption{Behavior of the density inferred from the \sii\ lines as a
function of the scale $L$ employed in different photoionization
calculations (each model is represented by a filled square). The
long-dashed and short-dashed line represent two dust-free sequences of
models in which the boundary density is $n_o = 150$ and 2000\cmc,
respectively. The dotted line and the solid line represent dusty models (all
with a boundary density $n_o = 150$) in which the $orion$ ($\mud=0.8$) or
the $noiro$ ($\mud=0.4$) extinction curve was employed, respectively.
} \label{fig:nsiiL}
\end{center}
\end{figure}

\section{Results of the models} \label{results}

Our main proposition is that only one basic variable of the nebular
geometry, the scale $L$, varies behind the slit aperture.  We will
show that the observations we have taken, albeit very limited, are
{\it not} inconsistent with this simple picture.

\subsection{Properties of a powerlaw stratification} \label{prop}

For a given dust content \mud\ and dust extinction curve, we proceed
to show that for a given value of $n_o$, there exists a unicity
relation between the \sii\ density (which is representative of the IF
electronic density) and the scale of the gradient $L$. This is shown
in Fig.~\ref{fig:nsiiL} where we plot the \sii\ density as a function of
the powerlaw scale $L$ employed in different photoionization
calculations.

In the dust-free case, the long-dashed line corresponds to the sequence
of models with $n_o = 150$\cmc\ while the short-dashed line
corresponds to models with a higher boundary density $n_o = 2000$\cmc.
For the denser model, the density gradient is much too shallow since
we get a $R_s$ of only 0.25\pc\ in the model with $L=0.7\pc$. Clearly,
the parameter $n_o$ plays an important role.  Its value can be deduced
by requiring that the best model sequence fits simultaneously the
observed profiles along the slit in \hb\ brightness and in \sii\
density.

\begin{figure} 
  \begin{center} \leavevmode
  \includegraphics[width=\columnwidth]{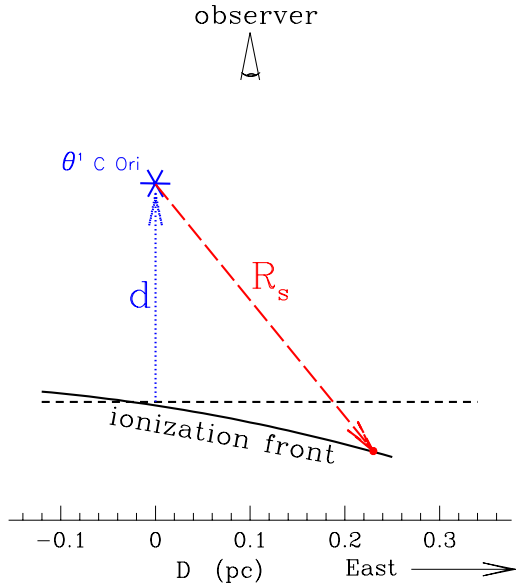}
\caption{Figure illustrating the sky projection of the West-East
aligned slit onto the nebular IF. The distance between \thetac\ and
the IF is $R_s$. The substellar distance is $d$ (see
Sect.~\ref{comp}). There is a further offset of the slit to the South
of \thetac\ of 0.073 pc (not shown).}
\label{fig:geom}
\end{center}
\end{figure}

In Fig.~\ref{fig:nsiiL}, the dotted line represents the case of using
the $orion$ extinction curve with \mud=0.8 (this sequence of models
have the same dust characteristics as those assumed by BF91).  The use
of the smaller grains $noiro$ curve (solid line) on the other hand
causes a considerable change in the inferred position of the IF even
though these models contain only {\it half} the mass of grains present in
the $orion$ curve models (dotted line). The reduction of
the IF radius is due to the amount of ionizing photons absorbed by the
dust grains.

In their determination of the IF shape, WO95 found a very interesting
albeit unexplained relationship between the distance to \thetac\ and
the gas density (their Fig.~6). They also found that the closer the IF
is to \thetac, the higher the \ha\ brightness appears to be. As shown
in Fig.~\ref{fig:nsiiL}, a powerlaw stratification establishes a
relationship between the scale $L$ and the \sii\ density near the
IF. A similar relationship holds between the \hb\ surface brightness
and the \sii\ density.  Our proposed powerlaw stratification is
therefore apt in providing a credible explanation to the intriguing
\sii\ density-radius correlation discovered by WO95.


\begin{figure}
  \begin{center} \leavevmode
  \includegraphics[width=\columnwidth]{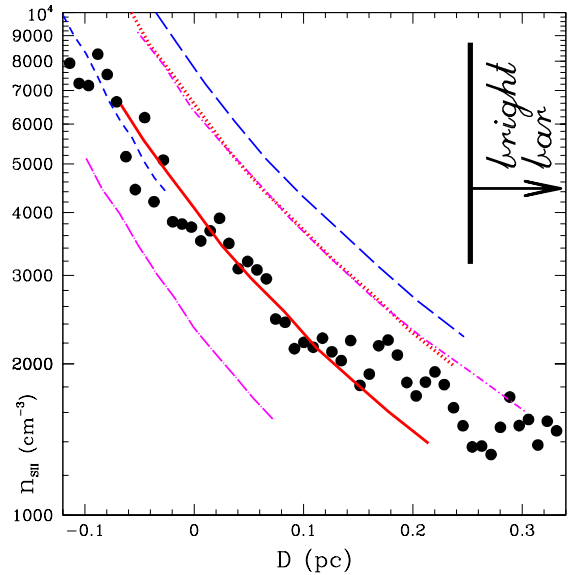}
\caption{Behavior of the \sii\ density after projection onto our
long-slit ($D$ axis).  The slit was aligned EW. $D=0$ corresponds to
the pixel nearest to \thetac\ with $D$ increasing towards the East,
crossing the Bright Bar near $D=0.25$. The long and short dash line
models represent the {\it dust-free} case with $n_o = 150$ and 2000\cmc,
respectively. The dotted and the solid lines represent models with
internal dust using the $orion$ and $noiro$ curves, respectively (both
with $n_o = 150$).  The two dot-dash lines represent the $noiro$ dusty
model using an boundary density of $n_o = 50$ (short-dash dotted line) and
500\cmc\ (long-dash dotted line), respectively. }  \label{fig:nsiiD}
\end{center}
\end{figure}

\begin{figure} 
  \begin{center} \leavevmode
  \includegraphics[width=\columnwidth]{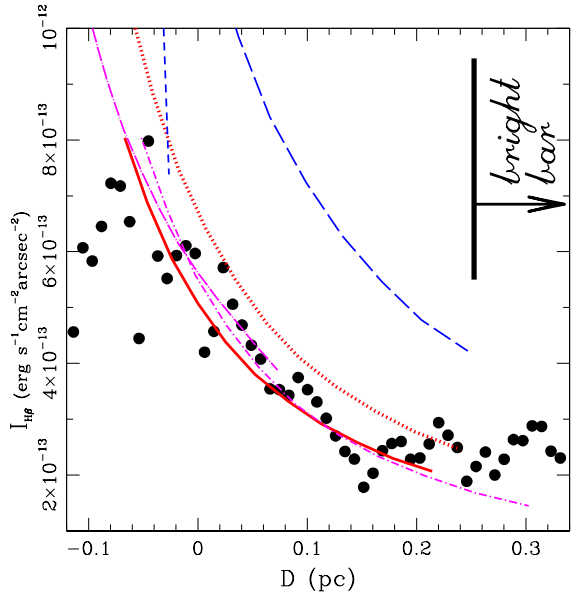}
\caption{Behavior of the \hb\ surface brightness in powerlaw density
stratified models after projection onto our long-slit.  The long-dash
and short-dash lines represent {\it dust-free} models with $n_o = 150$ and
2000\cmc, respectively. The dotted and solid line (both with $n_o =
150$\cmc) are models with internal dust using the $orion$ and $noiro$
curves, respectively. The two dot-dash lines represent the $noiro$
dusty model with boundary densities of $n_o = 50$ (short-dash dotted
line) and 500\cmc\ (long-dash dotted line), respectively. The observed
\hb\ surface brightnesses plotted here have been reduced to correct
for the contribution of the scattered light (Sect.\,\ref{dust}) and
for the contribution of the foreground component (Sect.\,\ref{geo}),
as in WO95 and BF91, respectively. } \label{fig:ihbD}
\end{center} 
\end{figure}

\subsection{Comparison with our long-slit data} \label{comp}

Comparison of the models with our long-slit data will be done
considering both the \sii\ density and the \hb\ surface
brightness. One difficulty resides in relating the local IF radius,
$R_s$, with position, $D$, along our slit. We have proceeded as
follows. We have not resolved the inverse problem of finding the
distance of the IF at each observed position along $D$.  Our main goal
consists rather in exploring whether a powerlaw density is generally
consistent with the data at hand.  With this perspective in mind, we
settled for the following simple relation between $D$ and $R_s$\,: $D
= \sqrt{R_s^2-0.073^2\pc} - d$ where 0.073\pc\ is the slit offset of
30\arcsec\ to the South, $R_s$ the true IF distance to \thetac\ for a
given slit position $D$ in parsec projected on the sky.  $D=0$
corresponds to the pixel nearest to \thetac\ with $D$ increasing
towards the East.  The additional parameter $d$ (in pc) was obtained
by iteration, using as constraint that our best model should fit as
closely as possible both the observed \sii\ densities and the \hb\
surface brightnesses.  If the IF lied on a perfect plane perpendicular
to our line-of-sight, $d$ would correspond to the substellar distance
(O'Dell 1994) and it would have to be subtracted in quadrature from
$r$ (as for the slit offset). This resulted, however, in an
unsatisfactory fit for all the models tried. A plausible explanation
can be found in the 3D-map of the IF shape of WO95 which shows that
the IF plane is tilted away from us at the corresponding position of
our slit. The above equation, in which $d$ is subtracted linearly,
qualitatively describes a tilted plane. A cartoon describing the
proposed IF geometry is shown in Fig.~\ref{fig:geom}.

In Fig.~\ref{fig:nsiiD}, we compare our stratified models with the \sii\
densities observed along the slit. For simplicity, all models cover
the same range in $L$, from 0.22 to 0.66\pc, which is the appropriate
range for the favored solid line model. This has the advantage of
clarifying the effect of varying only one parameter at a time (either
$n_o$ or \mud). The surface brightnesses along the slit are shown in
Fig.~\ref{fig:ihbD} and can be compared with the various model
sequences.  The two Figures were drawn using $d= 0.26$ which is the
value required by the favored model (solid line).

We emphasize that no dust-free model could provide an acceptable fit to
the data in both figures \ref{fig:nsiiD} and \ref{fig:ihbD}, whatever
value was used for $d$. In the case of dusty models using the
alternative extinction curve $orion$, the fit was never satisfactory
in both Figures simultaneously.  In effect, if we increase $d$ to
0.35, this shifts all model sequences to the left by 0.09\,pc in both
Figures, thereby resulting in a good fit of the $orion$ model in
Fig~\ref{fig:nsiiD} but at the expense of an unsatisfactory fit in
Fig~\ref{fig:ihbD}.  Increasing the dust content brings the dusty
models down but it is not possible in the case of the $orion$ since
$\mud = 0.8$ is already the maximum allowed given the subsolar
metallicity of Orion. The preeminence of small grains, on the other
hand, presents the advantage that less dust is necessary ($\mud = 0.4$
for the $noiro$ curve).  This scenario makes sense in the event that
dust grains have been progressively photoeroded.
 
In conclusion, even when allowing $d$ to vary freely, the Orion model
consisting of boundary density $n_o=150$\cmc\ and containing small
grains ($noiro$ with \mud = 0.4) provides the optimized set of
physical conditions\footnote{Interestingly, the apparent reddening
manifested in the emergent line ratios due to internal dust is very
small. In effect, the emergent Balmer decrement for the small-grain
and the dust-free models are 2.97 and 2.88 respectively} favored by
our model for the following reasons:
\begin{enumerate}

\item dust-free models are   unsuccessful in reproducing 
the observed behavior of the \hb\ surface brightness
of Fig.~\ref{fig:ihbD} (adopting   a larger value for $Q_H$ would
worsen the discrepancy),

\item for a given  dust content and size distribution, 
the determination of the boundary density $n_o$ is straightforward since
the curves using different densities lie relatively far apart in the
\sii\ density diagram of Fig.~\ref{fig:nsiiD} (for instance, compare the
position of the two dash-dotted lines corresponding to sequences with
boundary densities of $n_o=50$ or 500\cmc),

\item models favoring the preeminence of small grains (e.g. the $noiro$
extinction) provide a simultaneous fit to both data sets presented in
Figs~\ref{fig:nsiiD} and
\ref{fig:ihbD}, respectively.

\end{enumerate}

The applicability of the proposed model does not include the so-called
``Bright Bar'' ($\ge 0.25$ \pc) beyond which the
\hb\ brightness is slightly  rising. 
Accounting for this rise in our simple picture would require a gas
structure which lies closer to \thetac\ than the calculated IF model
above. Alternatively, the rise may be due to some sort of limb
brightening. Dopita et~al. (1974) have proposed for instance that the
Bright Bar correspond to the IF seen edge-on. This is qualitatively
corroborated by the IF distance map produced by WO95, which shows a
sharp ridge at the position of the Bar.

\begin{figure*} 
  \begin{center} \leavevmode
  \includegraphics[width=0.75\textwidth]{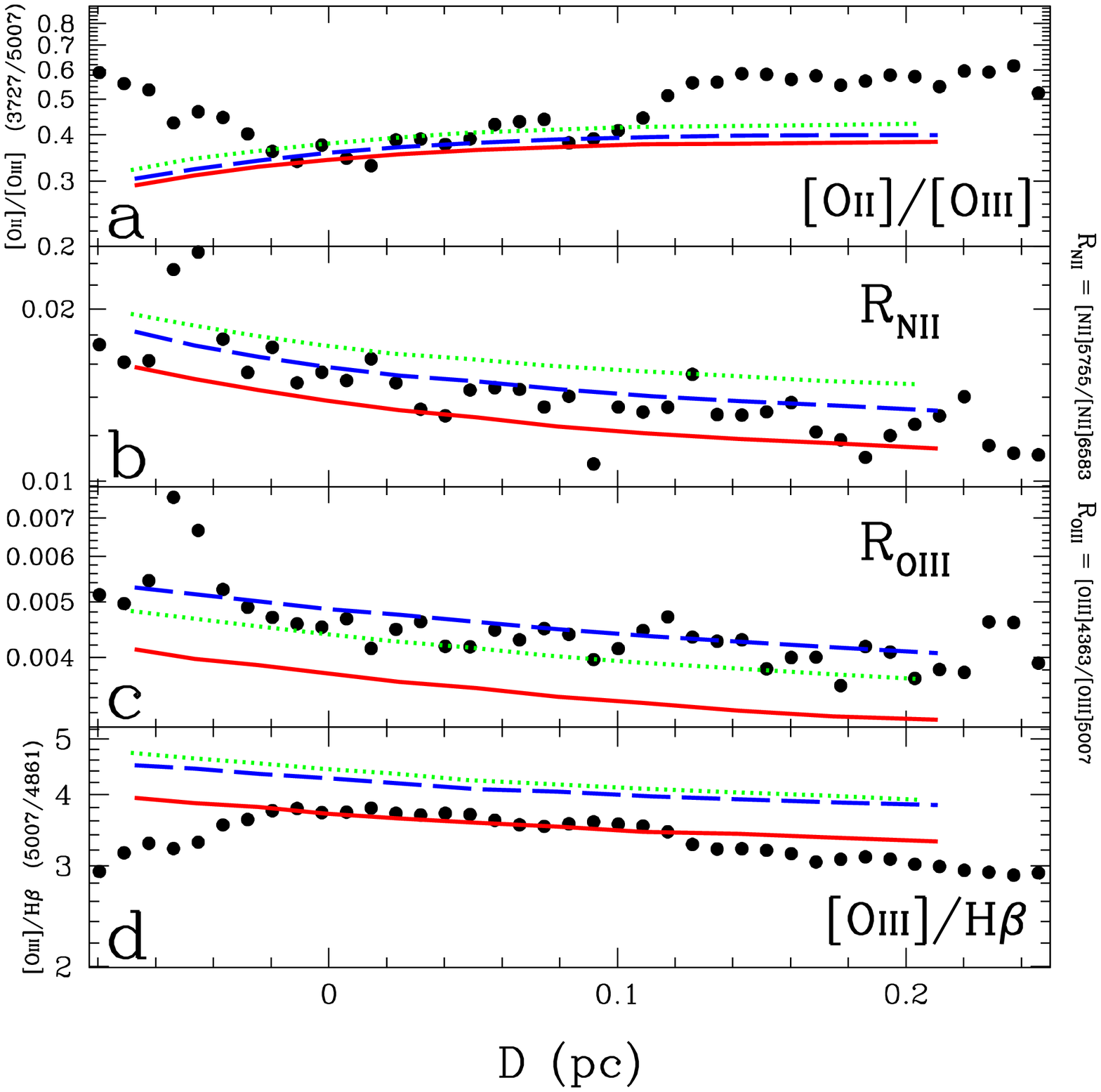}
\caption{The dots represent   the observed line ratios. 
Panel a: the line ratio \oiiw/\oiiiw\ which is a good indicator of the
degree of ionization, panel b: the Nitrogen temperature sensitive
ratio \niitw/\niiw, panel c: the Oxygen temperature sensitive ratio
\oiiitw/\oiiiw, panel d: the \oiiiw/\hbw\ ratio.   
The scale of the 3 panels b, c and d is the same and extend over 0.42
dex.  The models (see Sect.~\ref{lines}): the solid line shows the
behavior of line ratios along the slit for our most successful model
($n_o = 150$ and the $noiro$ curve with $\mud = 0.4$). The long-dashed
line was calculated assuming temperature fluctuations with $\tsq =
0.01$. The dotted line was calculated using half the diffuse field
from the ``outward-only'' approximation.  The spike in the data near
$D=-0.05$ pc in panels b and c is believed to be due to a foreground
proplyd.}
\label{fig:lines}
\end{center} 
\end{figure*}

\subsection{Gas temperature and line ratios  } \label{lines}

Although the current work does not attempt to provide a fit of all the
observed line ratios, it is worth checking whether the proposed
powerlaw density stratification results or not in ratios similar to the
observed. Because the ionization parameter in our models is not a free
parameter, line ratios become an interesting consistency
check. Furthermore, it is worth testing whether the models can
reproduce the observed line ratio trends with slit position.

In a similar fashion to BF91, one important parameter of our models is the
stellar temperature  which directly affects the excitation line ratio
\oiii/\hb. We found that reproducing satisfactorily this ratio
requires a stellar temperature of 38000 K when using the stellar
atmosphere models\footnote{More recent stellar atmosphere models exist
(Hubeny \& Lanz 1995; Schaerer \& de Koter 1997; Hillier \& Miller
1998; Pauldrach et al. 2001) but they differ considerably according to
whether they include or not the effect of stellar winds [c.f. Fig.~5
in Sankrit \& Hester (2000)]. Until the issue of wind loss is
resolved, all stellar temperatures inferred from photoionization
models will be of limited accuracy. The models of Hummer \& Mihalas
were used above for their convenience as they can be interpolated to
any arbitrary temperature within \map. Our results do not depend on
which family of models is used provided the stellar temperature is
adjusted to reproduce the line ratios well.} of Hummer \& Mihalas
(1970). This temperature is consistent with an evolved star of
spectral type O7 (c.f. WO95). Our line ratio calculations of
\oiii/\hb\ is represented by the solid line in panel d of
Fig.~\ref{fig:lines} using the same parameters as before, that is a
boundary density $n_o = 150$ and an extinction curve consisting of
small grains ($noiro$) with $\mud = 0.4$. We discuss below possible
explanations for the solid line lying slightly below the data. In any
event, it is encouraging that the models show the same gradient as the
data.

As for the temperature indicators $\rnii = \niitw/\niiw$ and $\roiii =
\oiiitw/\oiiiw$, a  gradient is also present in the data 
as shown by panels b and c (BF91 reported also a gradient of \rnii\
although for a different slit alignment). According to the models, the
gradient in the ratio \roiii\ is due to a temperature gradient
($\Delta T \simeq 500$ K) while in the case of Nitrogen it is due to
both a density gradient (via collisional deexcitation, c.f. BF91) and a
temperature gradient. Although the stratified models can successfully
fit the observed trends in panels b and c, the predicted ratios lie
somewhat lower than observed, more so for \roiii. Lowering the
metallicity would increase \roiii\ but at the expense of having the
\oiii/\hb\ ratio becoming larger than observed.

It is well known that photoionization models do not explain why the
temperatures derived from collisionally excited lines differ from that
of recombination lines (see review by Peimbert 1995). In the case of
Orion, Esteban et al. (1998) found that the abundance of the O$^{+2}$
ion derived from the \oiir\ recombination lines are incompatible with
the same ion abundance derived from the collisionally excited \oiiiw\
when a common temperature is adopted. Deriving a unique ionic
abundance was found to require different {\it mean} temperatures. These
can be justified if there exist high amplitude temperature
fluctuations in the nebula (Peimbert 1967) which far exceed the amplitude
arising from traditional photoionization models (Kingdon
\& Ferland 1995, 1998; P\'erez 1997). The mean square fluctuation
amplitudes that Esteban et al. inferred, lie in the range $\tsq =
0.02$--0.028. The effect of temperature fluctuations on line
emissivities were incorporated in the code \map\ along the scheme
described in Binette \& Luridiana (2000).  The long-dashed lines in
Fig.~\ref{fig:lines} represent such a model calculated with $\tsq =
0.01$. As expected, \rnii\ and, more so, \roiii\ are both
significantly increased and can now reproduce satisfactorily the data
of panels b and c. The calculated \oiii/\hb\ ratio in panel a,
however, lies now too high and the stellar temperature would have to
be reduced in order to produce a self-consistent model in all
panels. Given our current poor knowledge about the nature and cause of
the postulated temperature fluctuations, we have not carried out
further the iterative process.

The diffuse ionization radiation field plays a non-negligible impact
on the gas temperature. In our models, we use the ``outward-only''
approximation to treat the transfer of the diffuse radiation. This is
equivalent to assuming a single direction for the propagation of the
diffuse field, a technique based  on the approximation that the diffuse
field is soft and  does not therefore travel far  from the point of
emission. To test whether the integration along the direction of
increasing density may cause an overestimation of the intensity of the
diffuse radiation, we have calculated models in which we reduced by
half the strength of the diffuse field (adopting $\tsq = 0$). Such
models are represented by the dotted-line in Fig.~\ref{fig:lines}. We
can see a significant shift which is in the same direction as  that caused by
temperature fluctuations with $\tsq \sim 0.01$. The dotted line gives
us a useful upper limit on the line ratio uncertainties resulting from a
possible inadequacy of the outward-only approximation.

Panel a is a plot of the \oii/\oiii\ ratio. The models can reproduce
the observed points only within a restricted slit range.  No obvious
explanation can be found for the discrepancy outside it. It possibly
reflects the limitations of our simplified geometry as discussed in
Sect.~\ref{geo} or the larger errors affecting the UV \oii\
lines. Overall, the fit of the gradients using our simple models  
appears satisfactory in all the panels of Fig.~\ref{fig:lines}.

\section{Conclusions}

The method used by WO95 to determine the IF shape requires, at each
point of the nebula, knowledge of {\it both} the \ha\ surface
brightness and the IF density (using the \sii\ or \oii\ doublet). For
a powerlaw density stratification as proposed in this Paper, the
density across the nebula is not a free parameter since it is uniquely
determined by the local powerlaw scale $L$. This opens the possibility
that using only the \ha\ (or \hb) surface brightness map, one could in
principle determine the IF shape of the whole nebula. For this reason,
further studies are warranted in order to: $a)$ extend the comparison
of powerlaw density models to a more extensive set of observations of
Orion and $b)$
work out a scheme to solve for the inverse problem (as was done by
WO95 in the exponential case) in order to determine the IF shape
assuming the proposed powerlaw density gradient and taking into
account the effects of internal dust.

\acknowledgements 

LB and YDM acknowledge financial support from the CONACyT grants
32139-E and 25869-E, respectively.  We are thankful to M. Rodr\'\i guez
for sharing detailed information about her spectroscopic observations
of Orion. We also thank the unknown referee for the many constructive
comments made to the original draft.



\end{document}